\documentclass{webofc}
\usepackage[varg]{txfonts}   
\begin{document}
\title{Holographic spacetime from lattice Yang-Mills theory}
%
%

\author{\firstname{Niko} \lastname{Jokela}\inst{1,2}\fnsep\thanks{\email{niko.jokela@helsinki.fi}} \and
        \firstname{Arttu} \lastname{P\"onni}\inst{3}\fnsep\thanks{\email{arttuponni@gmail.com}} \and
        \firstname{Tobias} \lastname{Rindlisbacher}\inst{4}\fnsep\thanks{\email{trindlis@itp.unibe.ch}} \and
        \firstname{Kari} \lastname{Rummukainen}\inst{1,2}\fnsep\thanks{\email{kari.rummukainen@helsinki.fi}} \and
        \firstname{Ahmed} \lastname{Salami}\inst{1,2}\fnsep\thanks{\email{ahmed.salami@helsinki.fi}}
}

\institute{Helsinki Institute of Physics and Department of Physics, P.O. Box 64, FI-00014 University of Helsinki,
Finland
\and
           Micro and Quantum Systems Group, Department of Electronics and Nanoengineering, Aalto University, Finland
\and
           AEC, Institute for Theoretical Physics, University of Bern, Sidlerstrasse 5, CH-3012 Bern, Switzerland
          }

\abstract{%
Entanglement entropy is a notoriously difficult quantity to compute in strongly interacting gauge theories. Existing lattice replica methods have suffered from a severe signal-to-noise ratio problem,
making high-precision studies prohibitively expensive. Our improved lattice method mitigates this situation and allows us to probe holographic predictions for the behavior of entanglement
entropies in three- and four-dimensional Yang-Mills theories. We use this data for the numerical reconstruction of holographic bulk metrics.
}

\hfill HIP-2022-30/TH

\maketitle
%

\section{Introduction}\label{intro}

Color confinement in QCD is a very hard problem. Since the phenomenon is non-perturbative in nature, its description has defied theorists to date. There are heroic numerical works, but it is fair to say that we are still far from a complete picture. 

Instead of directly addressing confinement in QCD, one often also seeks simple models and aims to gain some understanding that could be transferred back to QCD. One possible route is offered via AdS/CFT or holography in short. In fact, one needs to break conformal symmetry in a controlled manner, a terminology better suited would be non-AdS/non-CFT correspondence. The holographic tools that are probes of confinement include various loops (Wilson, 't Hooft etc.), entanglement entropy (EE), and c-functions that count the number of degrees of freedom. In particular, entanglement c-functions seem quite intriguing, since they can also provide a clean score card on holography. We will mostly focus on this last probe.  

We are nowhere near solving confinement even in holographic large-$N$ field theories, but the hope is that the methods we build using lattice formulation will feed insight also into confinement in the long run. Our intermediate goals are rather challenging. They deal with also not-so-well-understood phenomena, entanglement of subsystems in gauge field theories, especially at strong coupling. The study of entanglement measures in gauge theories is of course interesting in its own right.

In this talk we reveal some results on our lattice exploration of entanglement entropies of bipartite pure systems in three- and four-dimensional bosonic pure glue Yang-Mills theories. As proxies for holographic suggestions for the behaviors of EEs (and Wilson loops) on what to expect, we consider supersymmetric versions of these field theories at large-$N$. We will be stupefied on how well the holographic expectations are met.

\section{Entanglement as a probe of confinement}\label{sec:EE}

Let us first recall the basic argument \cite{Nishioka:2006gr,Klebanov:2007ws} on why we expect some entanglement measure to indicate in which phase we are in at large-$N$. The idea is that in the deconfined phase of, say, SU($N$) gauge field theory, the degrees of freedom (dofs) that are effective are colorful ({\emph{e.g.}}, gluons) and the entropy density scales as $\sim \mathcal{O}(N^2)$ while in the confined phase the effective dofs are color singlets ({\emph{e.g.}}, glueballs) and the entropy density scales as $\sim \mathcal{O}(N^0)$. Now, if one is at large-$N$, in particular, a quantity that captures dofs (such as an entanglement entropic c-function) acts as a sharp order parameter. Indeed, this expectation has been met in holographic works \cite{Nishioka:2006gr,Klebanov:2007ws}, where the (derivative of the) holographic EE undergoes a transition from a finite to a vanishing value at around the scale of confinement. Later this argument has been sharpened to a statement \cite{Jokela:2020wgs} that the holographic EE undergoes this transition whenever there is an intrinsic energy scale in the system and so holographic EE probes the finite correlation length instead, beyond of which the EE is saturated. 

One goal in our program is to study entanglement c-function as extracted from lattice SU($N$) Yang-Mills theory and explore if a precursor to a sharp order parameter emerges with increasing $N$. The determination of entanglement measures in gauge theories is a difficult task, but with the so-called ``replica trick'', a family of entanglement measures, R\'enyi entropies, becomes available with lattice Monte Carlo. In this work we will essentially follow the prescription \cite{Buividovich:2008kq}. However, its direct implementation suffers from a severe signal-to-noise ratio problem, which makes accurate determination of R\'enyi entropies very expensive. The improved method we have developed \cite{Rindlisbacher:2022bhe} is powerful enough to make meaningful comparison to holographic predictions. In addition, we will provide astounding evidence that the finite part of the entanglement entropy for strip subsystems in three dimensions scales linearly with the width, the slope matching with the Bekenstein-Hawking entropy of the dual black brane.

\section{Bulk reconstruction}\label{sec:bulkreconstruction}

Once there is some data for a gauge field theory in question, one can ask whether there is a dual gravity description from where this data could in principle be obtained. This is an inverse problem, applied AdS/CFT done in reverse. The entanglement measure data is in some sense the optimal starting point: the holographic candidate \cite{Nishioka:2009un} only involves finding an extremal surface given a metric involving no assumptions on the matter content in the bulk. Of course, we are not able to construct all metric components. In particular, since our lattice method is naturally formulated in Euclidean signature, there are slim chances to get accurate predictions for time-dependent phenomena. Nevertheless, this limitation should not be taken as a discouragement, but perhaps be seen as an intermediate obstruction that can be circumvented via some analytic continuation to Lorentzian signature or with the help of bulk energy momentum tensor to fix $g_{tt}$, similarly to \cite{Saha:2018jjb}.

EE itself is not observable since it is UV-divergent. Some derived quantities are a better fit for bulk reconstruction. In \cite{Jokela:2020auu} it was proposed to use the derivative of the EE instead. Indeed, one can explicitly show that the derivative of EE with respect to the strip width, $\partial_\ell S_{EE}$, can be repackaged in a very convenient local expression with no explicit integrals, in terms of metric components evaluated at the bottom of the area functional whose extremal value captures the EE. The confidence intervals for the metric components are thereby fed from the statistical errors in the $\partial_\ell S_{EE}$ data in a very transparent manner.

\section{Results}\label{sec:results}

The writings of the articles are underway, but here we will present some results that illustrate how well the method of extracting the entanglement entropy derivative from the lattice works. We will mostly focus on three-dimensional SU(2) Yang-Mills theory, so the relevant quantity in question is $\partial S_{EE}/\partial \ell$ vs. $\ell$, where $\ell$ is the width of a strip subsystem. In three dimensions we mostly focus on finite temperature. 

\begin{figure}[htbp]
\centering
\includegraphics[width=0.8\linewidth]{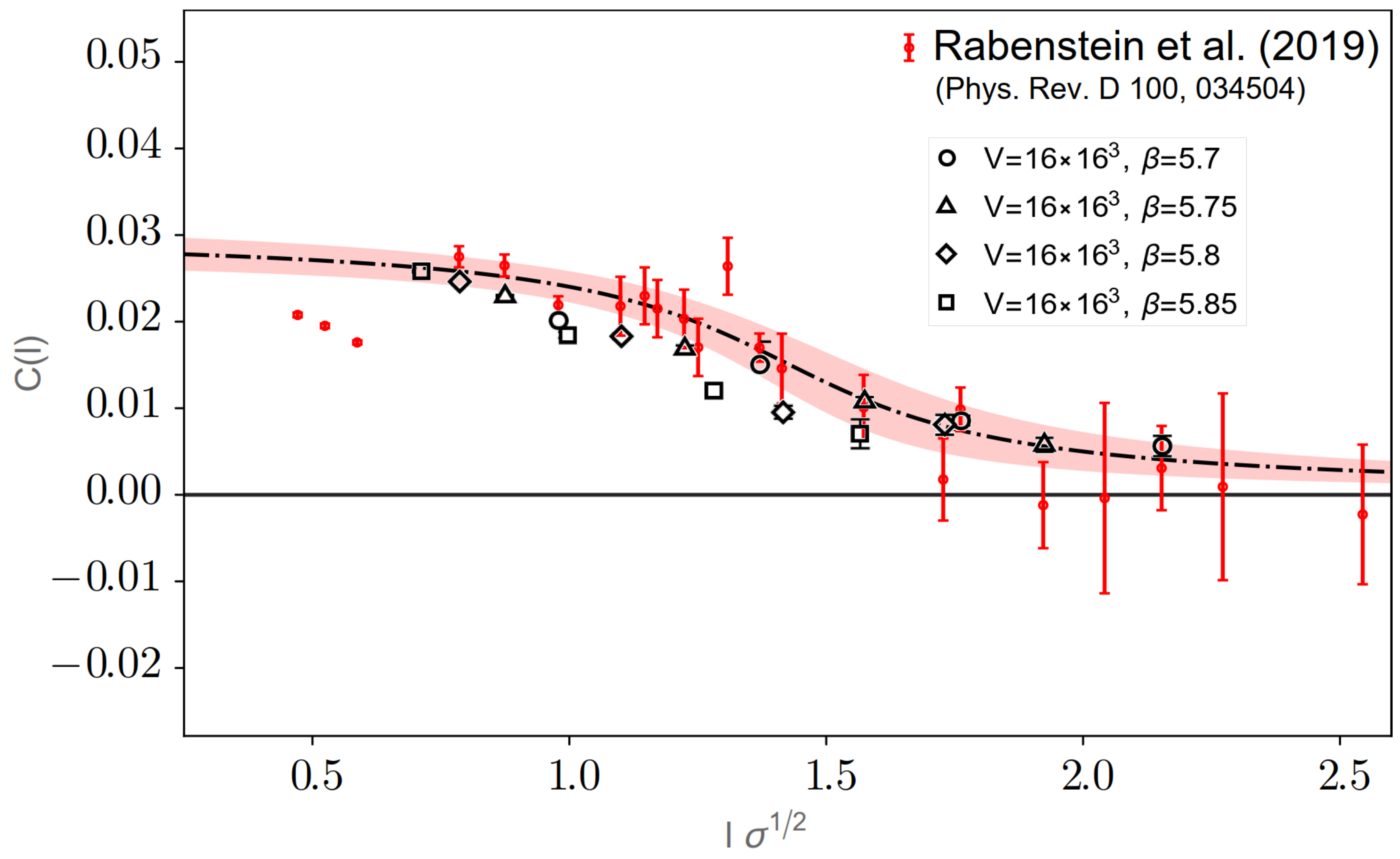}
\caption{Comparison of our c-function results (black markers) for SU$(3)$ in four dimensions with the results from \cite{Rabenstein:2018bri}. This is at vanishing temperature.}
\label{fig:cfunc4dc}
\end{figure}

In \cite{Rindlisbacher:2022bhe} and in the upcoming articles we also discuss four-dimensional YM theory with different numbers of colors in more depth. In particular, one interesting focal point is the entanglement c-function. In Fig.~\ref{fig:cfunc4dc} we depict preliminary results for the entanglement c-function $C(\ell)\propto \ell^3\partial_\ell S_{EE}$

\begin{figure*}[ht]
\centering
\includegraphics[width=0.8\textwidth,clip]{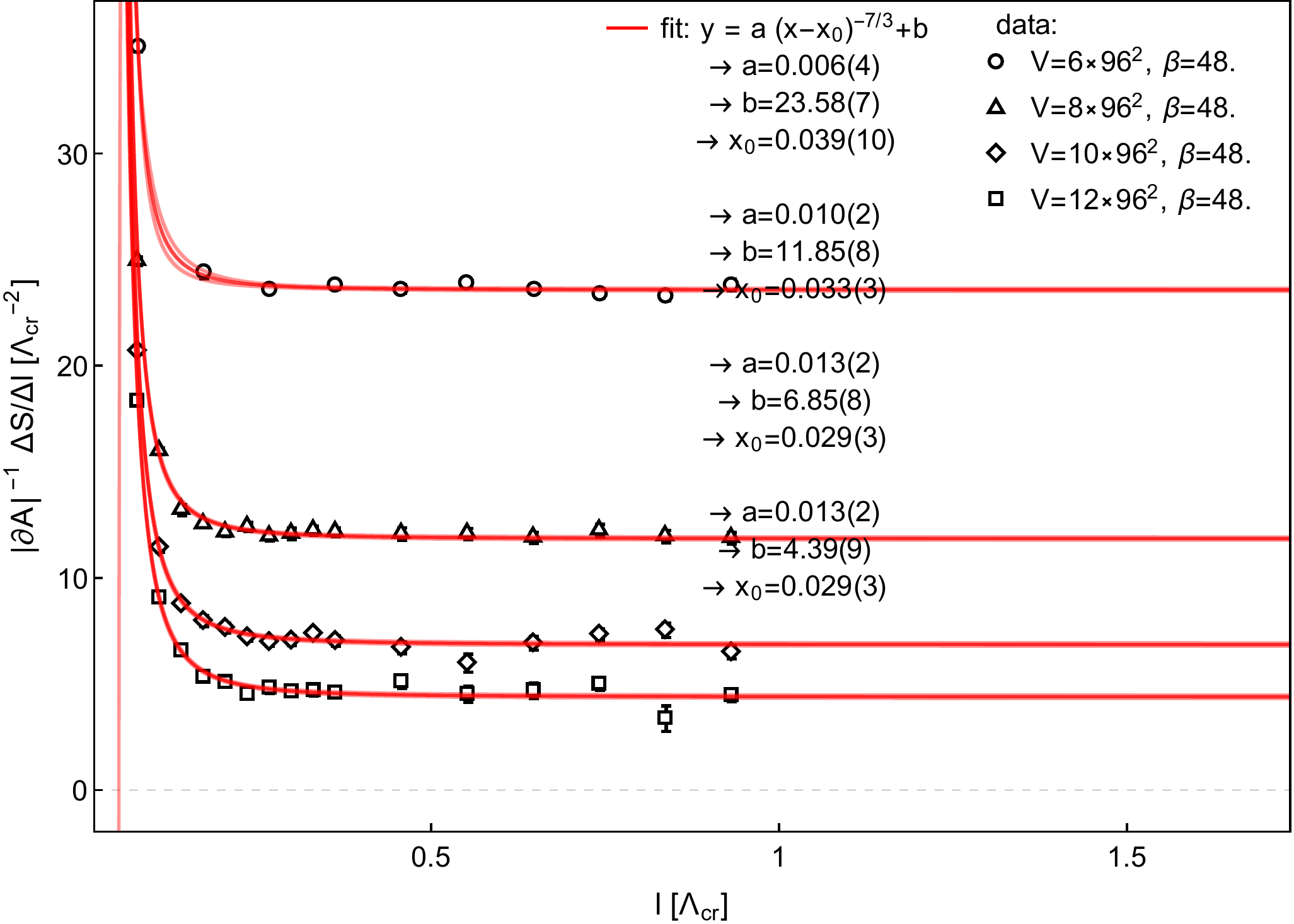}
\caption{The derivative of the entanglement entropy wrt. strip width $\ell$ in three-dimensional SU(2) Yang-Mills theory. The different curves represent increasing temperature (bottom-up). The fit here is just to guide the eye, more refined analyses reveal the IR (large-$\ell$) $\partial_\ell S_{EE}\sim T^{7/3}$ and UV (small-$\ell$) $\partial_\ell S_{EE}\sim \ell^{-7/3}$ behaviors. }
\label{fig:SEEvsl}       
\end{figure*}

Let us now continue to three dimensions. In Fig.~\ref{fig:SEEvsl} we depict the derivative of the entanglement entropy with respect to the strip width $\partial S_{EE}/\partial \ell$ for varying values of the temperature. We note two things. First of all, one finds that for small values of $\ell$, one gets $\partial_\ell S_{EE}\sim \ell^{-7/3}$; this exponent is fitted to good accuracy using a log-log scale. The exponent $-7/3$ is consistent with the expectation from D2-background, namely, that the finite part of EE scales as $S_{EE}\sim \ell^{-4/3}$ \cite{Maldacena:1998im}. 

The other observation is that $\partial_\ell S_{EE}$ saturates to a constant in the large-$\ell$ limit. This is also expected since the RT surface is supposed to trace the black hole horizon and the area functional should scale linearly with $\ell$. Indeed, we can associate this constant value with the Bekenstein-Hawking entropy of the black hole. We have, for example, checked that this scales as $T^{7/3}$, also expected for D2-brane background at finite temperature.

\begin{figure}[!ht]
\centering
\includegraphics[width=0.45\textwidth,clip]{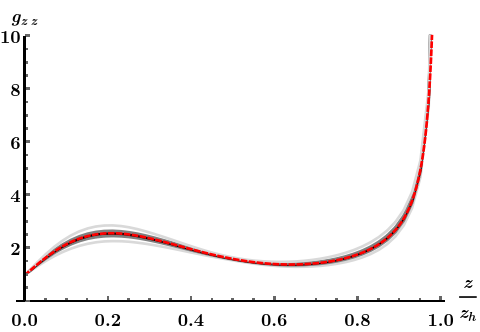}
\includegraphics[width=0.45\textwidth,clip]{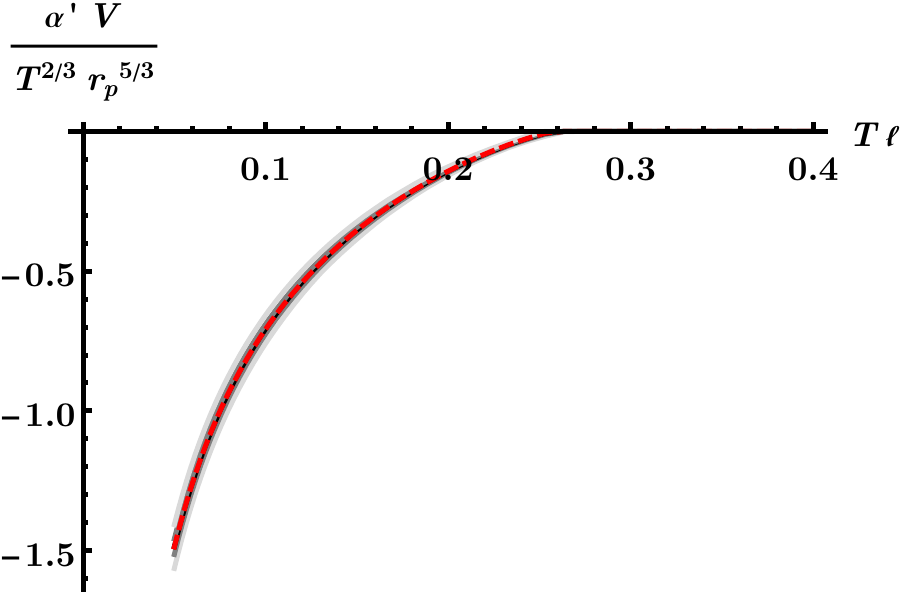}
\caption{{\bf{Left:}} One of the metric components as reconstructed from the SU$(2)$ lattice data for $\partial_\ell S_{EE}$ at $T/T_c=2.646$. {\bf{Right:}} The prediction for the static quark potential given the metric on the left. In these plots the dashed curve is the maximum likelihood estimate of the data. The black dashed curve is the median in the distribution of $g_{zz}$ while the dashed gray curves represent the $50\%$ and $95\%$ central confidence intervals, respectively.}
\label{fig:gzz}      
\end{figure}

Parts of the gravity metric can be readily reconstructed from the data, for example from that depicted in Fig.~\ref{fig:SEEvsl}. In Fig.~\ref{fig:gzz} we show radial-radial metric component of the metric, including the confidence intervals. This metric can then be used, in principle, to make some predictions for other quantities. Alas, not all metric components can be constructed this way, especially the time-time component is seemingly out of reach due to lattice bound to Euclidean form. In addition, the metric as constructed is in the Einstein frame, while for example the holographic prescription for computing the Wilson loop requires the string frame metric. Nevertheless, by assuming the D2-brane ansatz for the dilaton $e^\phi=(z/r_p)^{5/4}$, with $r_p$ the curvature radius, one obtains a reasonable result for the static quark potential; see Fig.~\ref{fig:gzz}. The potential vanishes at some critical separation, corresponding to fundamental string breaking transition in the bulk.

Encouraged by the unreasonable success of the holographic expectations and by especially the fact that the D2-brane background at finite temperature has its known limitations, we launched the computation of temporal Polyakov loops from the same lattice system. For small (but not too small) separation $\ell$ of the quark pair, the D2-brane background implies that the $q\bar q$-potential scales as $\ell^{-2/3}$ \cite{Maldacena:1998im}. The lattice results that we obtain meet this expectation. In the large separation limit the standard expectation is that the fundamental string breaks around the hadronic scale and one finds a flat potential as in Fig.~\ref{fig:gzz}. However, one can follow the prescription given in \cite{Albacete:2008dz}, {\emph{i.e.}}, ignore the string breaking and follow subdominant saddle for larger $\ell$ and finally to the complex plane. In the large-$\ell$ limit, we find $\text{Re} V\sim \ell^{-10/3}$ and $\text{Im}V\sim 1/\ell$. The data from the lattice we get is of course real but it is nevertheless consistent with the exponent $-10/3$. It is also consistent with the scaling that one expects from Debye screening $e^{-m_D \ell}/\sqrt{\ell}$ \cite{Dumitru:2002cf}. It would be interesting to dissect which one survives in the long run. For example, one could attempt to extract the imaginary part of the potential by following the real-time formalism in \cite{Burnier:2014ssa} and see if $1/\ell$ is obtained.

\section{Discussion}\label{sec:discussion}

We presented first results for the derivatives of the entanglement entropies for slab subsystems in bosonic SU($N$) Yang-Mills theories in three and four dimensions using our improved lattice method. We showed that many non-trivial holographic predictions are consistent to numerical accuracy with our data. We also showed that one can reconstruct the corresponding gravity dual metric, including confidence intervals. Lots of work is still needed in the construction of the dual geometry to large-$N$ QCD, but our work has led us to the era of precision holography. 

In addition to improving our understanding on what the holography has to offer to strongly coupled phases of gauge field theories, there are several other entanglement measures waiting to be calculated on the lattice. Among these are entanglement entropies for multiparty systems, but also mixed state entanglement entropies. Other interesting extensions would be towards anisotropy and address the murky situation of whether there even exists an entanglement monotone \cite{Hoyos:2021vhl}.

\paragraph{Acknowledgments}
N.~J. would like to thank the organizers of the XVth Quark Confinement and the Hadron Spectrum 2022 conference in Stavanger, Norway, for the invitation
to present this talk. The support of the Academy of Finland grants no. 1322307, 1320123, and 1345070 are acknowledged.

\bibliography{references}

\begin{thebibliography}{14}

\bibitem{Nishioka:2006gr}
T.~Nishioka, T.~Takayanagi, JHEP \textbf{01}, 090 (2007),
  \texttt{hep-th/0611035}

\bibitem{Klebanov:2007ws}
I.R. Klebanov, D.~Kutasov, A.~Murugan, Nucl. Phys. B \textbf{796}, 274 (2008),
  \texttt{0709.2140}

\bibitem{Jokela:2020wgs}
N.~Jokela, J.G. Subils, JHEP \textbf{02}, 147 (2021), \texttt{2010.09392}

\bibitem{Buividovich:2008kq}
P.V. Buividovich, M.I. Polikarpov, Nucl. Phys. B \textbf{802}, 458 (2008),
  \texttt{0802.4247}

\bibitem{Rindlisbacher:2022bhe}
T.~Rindlisbacher, N.~Jokela, A.~P\"onni, K.~Rummukainen, A.~Salami, PoS
  \textbf{LATTICE2022}, 031 (2022), \texttt{2211.00425}

\bibitem{Nishioka:2009un}
T.~Nishioka, S.~Ryu, T.~Takayanagi, J. Phys. A \textbf{42}, 504008 (2009),
  \texttt{0905.0932}

\bibitem{Saha:2018jjb}
A.~Saha, S.~Karar, S.~Gangopadhyay, Eur. Phys. J. Plus \textbf{135}, 132
  (2020), \texttt{1807.04646}

\bibitem{Jokela:2020auu}
N.~Jokela, A.~P\"onni, Phys. Rev. D \textbf{103}, 026010 (2021),
  \texttt{2007.00010}

\bibitem{Rabenstein:2018bri}
A.~Rabenstein, N.~Bodendorfer, A.~Sch{\"a}fer, P.~Buividovich, Phys. Rev.
  \textbf{D100}, 034504 (2019), \texttt{1812.04279}

\bibitem{Maldacena:1998im}
J.M. Maldacena, Phys. Rev. Lett. \textbf{80}, 4859 (1998),
  \texttt{hep-th/9803002}

\bibitem{Albacete:2008dz}
J.L. Albacete, Y.V. Kovchegov, A.~Taliotis, Phys. Rev. D \textbf{78}, 115007
  (2008), \texttt{0807.4747}

\bibitem{Dumitru:2002cf}
A.~Dumitru, R.D. Pisarski, Phys. Rev. D \textbf{66}, 096003 (2002),
  \texttt{hep-ph/0204223}

\bibitem{Burnier:2014ssa}
Y.~Burnier, O.~Kaczmarek, A.~Rothkopf, Phys. Rev. Lett. \textbf{114}, 082001
  (2015), \texttt{1410.2546}

\bibitem{Hoyos:2021vhl}
C.~Hoyos, N.~Jokela, J.M. Pen\'\i{}n, A.V. Ramallo, J.~Tarr\'\i{}o, JHEP
  \textbf{10}, 112 (2021), \texttt{2104.11749}

\end{thebibliography}

\end{document}